\documentclass[12pt]{article}
\usepackage{epsfig}

\newcommand\pubnumber{SLAC--PUB--11154}
\newcommand\pubdate{\today}
\newcommand\hepnumber{hep-ph/0505037}

\def\SLAC{Stanford Linear Accelerator Center\\
    Stanford University, Stanford, California 94309, USA}
\def\doeack{\footnote{Work supported in part by the Department of Energy,
                     contract DE--AC02--76SF00515.}}

\def\Title#1{\begin{center} {\Large #1 } \end{center}}
\def\Author#1{\begin{center}{ #1} \end{center}}
\def\Address#1{\begin{center}{ \it #1} \end{center}}

\newcommand\pubblock{\rightline{\begin{tabular}{l} \pubnumber\\
         \pubdate \\ \hepnumber \end{tabular}}}
\newenvironment{Abstract}{\begin{quotation} \begin{center}
                       ABSTRACT
     \end{center}\bigskip  }{\end{quotation}}

\def \as {\relax\ifmmode\alpha_s\else{$\alpha_s${ }}\fi}
\def \LQCD {\Lambda_{\mbox{\tiny QCD}}}
\def \O {\Omega}

\begin{document}
\begin{titlepage}
\pubblock

\vfill \Title{\bf Dijet Event Shapes as Diagnostic Tools } \vfill
\Author{ Carola F.~Berger\doeack} \Address{\SLAC}  \vfill
\begin{Abstract}
Event shapes have long been used to extract information about hadronic
final states and the properties of QCD, such as particle spin and
the running coupling. Recently, a family of event shapes,
the \emph{angularities}, has been introduced that depends on a continuous
parameter. This additional parameter-dependence further
extends the versatility of event shapes. It provides a handle on
nonperturbative power corrections, on non-global logarithms, and on
the flow of color in the final state.
\end{Abstract}
\vfill \vfill
\end{titlepage}
\def\thefootnote{\fnsymbol{footnote}}
\setcounter{footnote}{0}

\section{Introduction}

Event shapes are generalizations of jet cross sections that
describe the distribution of radiation in the final state. Because
no individual hadrons are observed, event shapes are
infrared-safe. This renders the main features of event shapes,
which are determined by the underlying hard scattering,
perturbatively calculable (for recent summaries see
Refs.~\cite{Dasgupta:2003iq,Banfi:2004nk}). Event shape
observables are therefore an ideal testing ground of QCD. For
example, they allow the determination of the gluon
spin~\cite{DeRujula:1978yh}, of QCD color
factors~\cite{Heister:2002tq}, and precise measurements of the
running coupling~\cite{Bethke:2004uy}.

Nevertheless, event shape observables retain sensitivity to
long-distance hadronization effects. For average values of
$e^+e^-$ event shapes, these nonperturbative corrections manifest
themselves as more-or-less additive corrections, typically
proportional to the inverse power of the center-of-mass (c.m.)
energy, $1/Q$. Differential distributions, which give important
additional information about the dynamics of QCD, however, require
a more involved convolution with functions that parameterize
hadronization effects (for reviews see for example
Refs.~\cite{Dasgupta:2003iq,Magnea:2002xt}).

Below we will illustrate how the introduction of a
parameter-dependence into an event shape observable can
considerably extend the information that can be extracted from its
measurement. We discuss a set of event shapes, the so-called class
of
\emph{angularities}~\cite{Berger:2002ig,Berger:2003iw,Berger:2004xf},
that depends on a continuous real parameter. This
parameter-dependence allows to test nonperturbative features of
event shape distributions~\cite{Berger:2003pk}, to control large
non-global logarithmic
corrections~\cite{Berger:2002ig,Berger:2003iw}, and, in the case
of hadronic collisions, to obtain information about the underlying
color flow~\cite{Berger:2001ns}.

We begin with a necessarily incomplete review of a prominent event
shape, the thrust~\cite{Brandt:1964sa,Farhi:1977sg}, and introduce
the family of angularities. We then derive the scaling rule for
nonperturbative power corrections to distributions of
angularities. Next, we illustrate how the correlation of
angularities with non-global observables, that is, with
observables that are designed to measure radiation into only part
of phase space, can be used to dampen the sensitivity of the
perturbative calculation to soft-gluon radiation at wide angles
away from the point of interest. In Section~\ref{color} we propose
an extension of the definition of angularities to hadronic
collisions, and show how such an observable may be used to extract
information about the distribution of color in the underlying hard
scattering. We end with a summary and point out future directions.

\section{The Thrust and Its Extension to the Family of Angularities}

\subsection{Determination of Gluon Spin via the Thrust}

One of the prime examples of an event shape observable in $e^+e^-$
collisions is the thrust~\cite{Brandt:1964sa,Farhi:1977sg} T:
\begin{equation}
T = \frac{1}{\sum\limits_{j \in N} \left| \vec{p}_j \right|}
\max_{\hat{n}} \sum\limits_{i \in N} \left| \vec{p}_i \cdot
\hat{n} \right|  \label{thrusteq1} \, ,
\end{equation}
where the sums are over all particles in the final state $N$, and
$\hat{n}$ is a unit vector whose direction is called the thrust axis
when $T$ is maximal. For massless partons, this definition can be
written equivalently as
\begin{equation}
\tau \equiv 1-T = \frac{1}{Q} \sum_{i \in N}\ p_{\perp,i }\;
e^{-|\eta_i|} \, ,\label{thrusteq2}
\end{equation}
where $p_{\perp,i}$ and $\eta_i$ denote the transverse momentum
and pseudorapidity $\eta_i = \ln \cot (\theta_i/2)$, respectively,
of particle $i$, with respect to the thrust axis. The maximization
is implied. Here and below, we treat all partons as massless,
parton masses induce corrections which have been examined in
Ref.~\cite{Salam:2001bd}.

The thrust measures how pencil-like a particular event is. In the limit
of two back-to-back jets, the value of $T$ is 1, whereas its
minimum value of 1/2 corresponds to a completely homogeneous spherical
event. Radiative corrections shift the value of the lowest order cross
section away from $T = 1$. The distribution of radiation, and thus
the differential cross section  $d\sigma/dT$, depends on the spin of the
radiated partons. The first-order QCD cross section is given by
\begin{equation}
\frac{1}{\sigma_{\mbox{\tiny tot}}} \frac{d \sigma(q\bar{q}g)}{d T} =
\frac{2\alpha_s}{3 \pi}
\Bigg[ \frac{2 (3 T^2 - 3 T+2)}{T(1-T)} \ln\left(\frac{2T-1}{1-T}\right)
 - \frac{3(3T-2)(2-T)}{1-T} \Bigg]\, , \label{thrustas1}
 \end{equation}
whereas the corresponding cross section where a scalar ``gluon''
is radiated gives a distinctly different distribution,
\begin{equation}
\frac{1}{\sigma_{\mbox{\tiny tot}}} \frac{d \sigma(q\bar{q}S)}{d T} =
\frac{\tilde{\alpha}_s}{3 \pi}
\Bigg[ 2 \ln\left(\frac{2T-1}{1-T}\right)
 + \frac{(4-3T)(3T-2)}{1-T} \Bigg]\, . \label{thrustscal}
 \end{equation}
This sensitivity to the spin of the radiated partons was used as
one of the first tests of QCD~\cite{DeRujula:1978yh}. The
definitions of a multitude of other $e^+e^-$ event shapes
measuring various aspects of angular distributions in the final
state can be found, for example, in Ref.~\cite{Dasgupta:2003iq}.

\subsection{The Family of Angularities}

We now extend the definition of the thrust,
Eq.~(\ref{thrustscal}), by introducing a dependence on a
continuous, real parameter $a$, which allows us to study a whole
class of event shapes simultaneously. We define the family of
angularities by weighing the final state with the
function~\cite{Berger:2002ig,Berger:2003iw,Berger:2004xf}
\begin{equation}
\tau_a(N) = {1\over Q}\sum_{i \in N}\ p_{\perp,i }\;
e^{-|\eta_i|{ (1-a)}}\,   \label{eventdef}\, ,
\end{equation}
where the direction of particle $i$ is again measured relative to
the thrust axis. The parameter $a$ varies between $a = 2$ (the
weight function (\ref{eventdef}) then selects two infinitely
narrow, back-to-back jets) and  $a = -\infty$ (the fully inclusive
limit). The value $a = 0$ corresponds to $1-T$, with $T$ the
thrust (\ref{thrusteq1}), and $a = 1$ corresponds to the jet
broadening~\cite{Catani:1992jc}.

In the two-jet limit, $\tau_a$ approaches $0$, and the
differential cross section $d\sigma/d\tau_a$ receives large
corrections in $\ln \left(1/\tau_a\right)$ due to soft gluon
radiation which have to be resummed for a reliable theoretical
prediction.  This resummation has been performed to all
logarithmic orders at leading power for $a < 1$ in
Ref.~\cite{Berger:2003iw}. For $a \sim 1$, recoil effects have to
be taken into account, as was pointed out for the broadening ($a =
1$) in Ref.~\cite{Dokshitzer:1998kz}. All equations below are
therefore valid in the range $a < 1$, where the value of $a$ is
not too close to 1.

We quote the result of the resummation of large
logarithms of $\tau_a$ in Laplace moment space:
\begin{equation}
\tilde{\sigma} \left(\nu,Q,a \right) =   \int^\infty_0 d \tau_a\, {\rm e}^{\;
-\nu\,
\tau_a}\ {d
\sigma(\tau_a,Q) \over d \tau_a}\, .
\label{trafo}
\end{equation}
Logarithms of $1/\tau_a$ are transformed into logarithms of $\nu$.
\begin{samepage}
At next-to-leading logarithmic (NLL) order, we find
\begin{eqnarray}
 {1\over \sigma_{\rm tot}} \, \tilde{\sigma}
\left(\nu,Q,a \right) &
= &
      \exp \Bigg\{ 2\, \int\limits_0^1 \frac{d u}{u} \Bigg[ \,
      \int\limits_{u^2 Q^2}^{u Q^2} \frac{d p_\perp^2}{p_\perp^2}
A\left(\as(p_\perp)\right) \left( {\rm e}^{- u^{1-a}
\nu \left(p_\perp/Q\right)^{a} }-1 \right)
\nonumber \\
 &  & \qquad \quad \qquad \qquad  + \,\frac{1}{2}
 B\left(\as(\sqrt{u} Q)\right) \left( {\rm e}^{-u
\left(\nu/2\right)^{2/(2-a)} } -1 \right)
      \Bigg] \Bigg\}.
\label{thrustcomp}
\end{eqnarray}
\end{samepage}
We see that for small $\tau_a$, the cross section factorizes into
contributions from two independently radiating jets, resulting in
the Sudakov exponent above, where the factor of two stems from the
two equal jet contributions. Here, $A(\alpha_s)$ and $B(\alpha_s)$
are well-known anomalous dimensions that are independent of $a$.
They have finite expansions in the running coupling, $A(\as)=
\sum_{n=1}^\infty A^{(n)}\ (\as/ \pi)^n$, and similarly for
$B(\as)$, with the coefficients  $A^{(1)} = C_F , \, B^{(1)}  =
-3/2 \, C_F ,\,  A^{(2)} = 1/2 \,C_F [ C_A ( 67/18 - \pi^2/6) -
10/9\, T_F N_f ]$, at NLL accuracy, with the color factors $C_F =
4/3$, $C_A = 3$, $T_F = 1/2$, and $N_f$ denotes the number of
flavors. At $a = 0$ we reproduce the NLL resummed thrust cross
section~\cite{Catani:1992ua}. The explicit NLL expression in
transform space can be found in Ref.~\cite{Berger:2003pk}, and a
formula for the resummed cross section valid to all logarithmic
orders at leading power is given in Ref.~\cite{Berger:2003iw}.

\section{Universality of Power Corrections}

As already mentioned in the introduction, the resummed expression
(\ref{thrustcomp}) is plagued by sensitivity to long-distance
effects. This sensitivity manifests itself as an ambiguity in how
to deal with the not well-defined integral over the running
coupling, and the validity of the perturbative approach breaks
down at $\tau_a \sim \LQCD/Q$. There exist a variety of approaches
(see for example Ref.~\cite{Magnea:2002xt}), but in all cases one
has to supplement the perturbative calculation with additional
information. In the present case, as we will see below, the
knowledge of nonperturbative (NP) corrections for only one
specific value of $a$ suffices to determine unambiguously the full
distribution, including power corrections, for all other values of
$a$ ($a < 1$).

Due to the quantum mechanical incoherence of short- and
long-distance effects, we can separate the perturbative part from
the NP contribution in a well-defined, although
prescription-dependent manner. Following
Refs.~\cite{Korchemsky:1998ev,Korchemsky:1999kt}, we deduce the
structure of the NP corrections by a direct expansion of the
integrand in the exponent at momentum scales below an infrared
factorization scale $\kappa$. We rewrite
 Eq.~(\ref{thrustcomp}) as the sum of a perturbative term,
 labelled with the subscript PT,
 where all $p_\perp > \kappa$, and a soft term that contains all NP physics:
\begin{eqnarray}
\ln \left[ \frac{1}{\sigma_{\mbox{\tiny tot}}} \tilde{\sigma} \left(\nu,Q,a \right) \right]
& = & \!\! 2 \left[\int\limits_{\kappa^2}^{Q^2}  \frac{d p_\perp^2}{p_\perp^2} +
\int\limits_{0}^{\kappa^2} \frac{d p_\perp^2}{p_\perp^2} \right]  A\left(\as(p_\perp)\right)
\!\!\!\!\! \int\limits_{p_\perp^2/Q^2}^{p_\perp/Q} \!\!
\frac{d u}{u}    \left( e^{- u^{1-a} \nu \left(p_\perp/Q\right)^{a} }-1 \right) \nonumber \\
& & \qquad
   + \, B \mbox{-term}
 \nonumber \\
 & \equiv & \ln \left[ \frac{1}{\sigma_{\mbox{\tiny tot}}} \tilde{\sigma}_{\mbox{\tiny PT}}
 \left(\nu,Q,\kappa,a \right) \right] \nonumber \\
 & & \qquad \qquad + \frac{2}{1-a} \sum\limits_{n=1}^\infty \frac{1}{n\, n!}
 \left(-\frac{\nu}{Q}\right)^n
\int\limits_{0}^{\kappa^2} \frac{d p_\perp^2}{p_\perp^2} p_\perp^n  A\left(\as(p_\perp)\right).
\label{nonpt}
\end{eqnarray}
We have suppressed terms of order ${\mathcal{O}}(\nu/Q^{2-a} , \nu^{\frac{2}{2-a}}/Q^2 )$,
 which include the entire $B$-term of Eq. (\ref{thrustcomp}), as indicated.
 Introducing the shape function as an expansion in powers of $\nu/Q$,
 from the expansion of the exponent in the second equality of (\ref{nonpt}),
 with NP coefficients $\lambda_n(\kappa)$,  we arrive at,
\begin{eqnarray}
\tilde{\sigma} \left(\nu,Q,a \right)
& = & \tilde{\sigma}_{\mbox{\tiny PT}} \left(\nu,Q,\kappa,a \right)\,
\tilde{f}_{a,\mbox{\tiny NP}} \left(\frac{\nu}{Q},\kappa\right) , \,\,\,\label{ptnp} \\
\ln \tilde{f}_{a,\mbox{\tiny NP}} \left(\frac{\nu}{Q},\kappa\right) &
\equiv & \frac{1}{1-a} \sum\limits_{n=1}^\infty \lambda_n(\kappa) \left(-\frac{\nu}{Q}\right)^n.
\end{eqnarray}
We find the simple result that the only dependence on $a$ is
through an overall factor $1/(1-a)$ which leads to the scaling
rule for the shape function~\cite{Berger:2003pk}:
\begin{equation}
\tilde{f}_{a,\mbox{\tiny NP}} \left(\frac{\nu}{Q},\kappa\right)  =
\left[ \tilde{f}_{0,\mbox{\tiny NP}} \left(\frac{\nu}{Q},\kappa\right)
\right]^{\frac{1}{1-a}}. \label{rule}
\end{equation}
For example, given the shape function  for
the thrust, $a = 0$, at a specific c.m. energy $Q$,
one can predict the shape function and thus from Eq.~(\ref{ptnp})
 the complete cross section including all leading power
 corrections for any other value of $a$.

\begin{figure}[h]
\centerline{\psfig{file=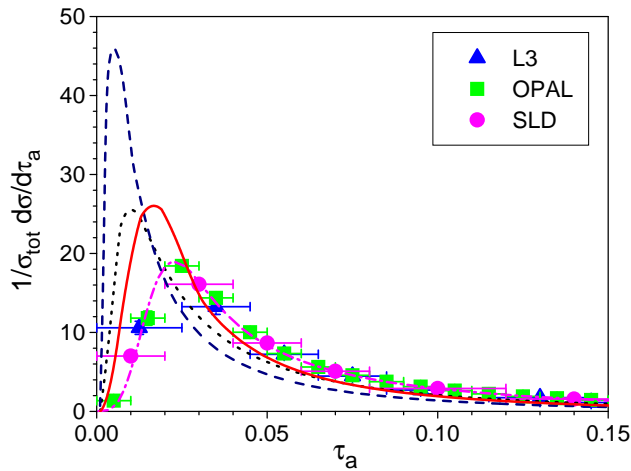,height=8.5cm,angle=270,clip=0}}
\vspace*{8pt} \caption{Differential distributions
$(1/\sigma_{\mbox{\tiny tot}}) d \sigma/d \tau_a$ for $a = 0$, and
$a = -0.5$ at $Q = 91$ GeV. Dash-dotted line: output of PYTHIA, $a
= 0$; dotted line: perturbative contribution at NLL/NLO, $a = 0$,
as defined in Eq.~(\ref{nonpt}); dashed line: same as dotted line,
at $a = -0.5$; solid line: prediction for $a = -0.5$ using Eqs.
(\ref{ptnp}) and (\ref{rule}). The data are taken from
Refs.~\protect\cite{Adeva:1992gv,Acton:1993zh,Abe:1994mf}.}
\label{figscale}
\end{figure}

This is illustrated in Fig.~\ref{figscale}, where we plot the
differential distributions for the thrust ($a = 0$) and for
angularity $a = -0.5$. The dotted ($a = 0$) and dashed ($a =
-0.5$) lines are the theoretical predictions at NLL from
Eq.~(\ref{thrustcomp}) matched to fixed next-to-leading order
(NLO) calculations with EVENT2~\cite{Catani:1996jh}. In order to
compute the shape function via Eq.~(\ref{ptnp}) for $a = 0$, we
use PYTHIA's~\cite{Sjostrand:2000wi} output at $a = 0$
(dash-dotted line), instead of fitting a function to the thrust
data, since PYTHIA fits these data very well. The prediction for
the full differential distribution for $a = -0.5$ (solid line) is
then found via (\ref{ptnp}) and (\ref{rule}) from the so
determined shape function and the NLL/NLO computation (dashed
line).

The above derivation rests on two main assumptions: Our starting
point, the NLL resummed cross section (\ref{thrustcomp}),
describes independent radiation from the two primary outgoing
partons. Inter-hemisphere correlations are neglected, although
they are present in the resummed formula valid to all logarithmic
orders~\cite{Berger:2003iw}. However, it has been found from
numerical studies that such correlations may indeed be
unimportant~\cite{Korchemsky:2000kp,Belitsky:2001ij,Gardi:2002bg}.
Furthermore, we assume in the separation of perturbative and NP
effects (\ref{nonpt}), that long-distance physics has the same
properties under boosts as the short-distance radiation. Success
or failure of the scaling (\ref{rule}) would thus provide
information about these properties of long-distance dynamics.

\section{Non-Global Logarithms}

The parameter-dependence of the family of angularities can also help
to control the sensitivity to wide-angle soft gluon radiation
of certain less inclusive, but still infrared
safe, so-called non-global, observables.

\subsection{Non-Global Observables}

\begin{figure}[htb]
\begin{center}
\epsfig{file=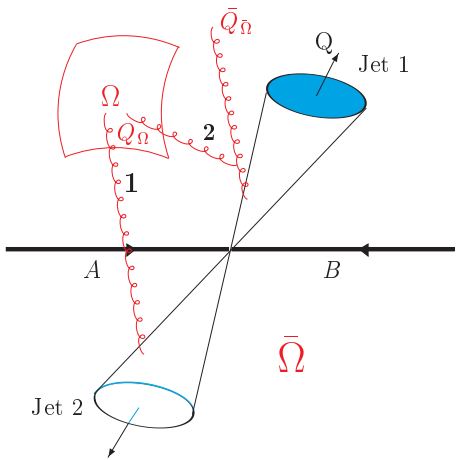,height=6.6cm,clip=0}
\caption{Sources of global and non-global logarithms in dijet
events. Configuration 1, a primary emission, is the source of
global logarithms, configuration 2 results in non-global
logarithms.} \label{fignonglobal}
\end{center}
\end{figure}

Non-global observables measure the radiation into only part of
phase space, an example is the energy flow into interjet
regions~\cite{Berger:2001ns,Sveshnikov:1995vi}. However, these
observables retain sensitivity to radiation away from the location
of interest via secondary radiation, as shown in
Fig.~\ref{fignonglobal}~\cite{Dasgupta:2001sh,Dasgupta:2002bw,Appleby:2002ke}.
Fig.~\ref{fignonglobal} illustrates an energy flow observable
associated with radiation into a chosen interjet angular region,
$\O$ with the complement of $\O$ denoted by $\bar\O$. We are
interested in the distribution of $Q_\O$ for events with a fixed
number of jets in $\bar \O$,
\begin{equation}
A + B \rightarrow \mbox{  Jets }  + X_{\bar{\O}}
   + R_\O (Q_\O)\, .
\label{event}
\end{equation}
Here $X_{\bar\O}$ stands for radiation into the regions
between $\O$ and the jet axes, and $R_\O$ for
radiation into $\O$.

As shown in Fig.~\ref{fignonglobal}, for these kinds of
observables there are two main sources of large logarithmic
corrections: ``primary'' emissions, such as gluon 1 in Fig.\
\ref{fignonglobal}, are emitted directly from the hard partons
into $\O$. Phase space integrals for these emissions contribute
single logarithms per loop, of order $\as^n\ln^n (Q/Q_\O) \equiv
\as^n \ln^n 1/\varepsilon$, where we have introduced the variable
$\varepsilon = Q_\O/Q$.  These logarithms exponentiate as above,
(\ref{thrustcomp}), and may be resummed in a straightforward
fashion~\cite{Berger:2002ig}.  There are also ``secondary"
emissions originating from the complementary region $\bar{\O}$,
illustrated by configuration 2 in Fig.\ \ref{fignonglobal}. As
emphasized in Refs.~\cite{Dasgupta:2001sh,Dasgupta:2002bw},
emissions into $\O$ from such secondary partons can give
logarithms of the form $\as^n \ln^n (\bar{Q}_{\bar{\O}}/Q_\O)$,
where $\bar{Q}_{\bar{\O}}$ is the maximum energy of radiation in
$\bar{\O}$.  These have become known as non-global logarithms. If
no restriction is placed on the radiation into $\bar{\O}$, then
$\bar{Q}_{\bar{\O}}$ can approach $Q$, and the non-global,
secondary logarithms can become as important as the primary
logarithms.  The non-global logarithms arise because real and
virtual enhancements associated with secondary emissions do not
cancel each other fully at fixed $Q_\O$, or equivalently
$\varepsilon$.

Such non-global logarithms cannot be factorized into a fixed
number of jets as in Eq.~(\ref{thrustcomp}), and therefore do not
exponentiate straightforwardly. However, it was found in
Ref.~\cite{Banfi:2002hw} that in the limit of large numbers of
colors the radiative corrections to non-global energy flow
observables
 can be described by
a non-linear evolution equation, which is formally equal to the
Kovchegov equation that describes the high-energy behavior of the
$S$-matrix~\cite{Kovchegov:1999yj}. Moreover, for a specific
non-global observable, the heavy-quark multiplicity in a certain
region of phase space,
 the non-global evolution
linearizes~\cite{Marchesini:2003nh,Marchesini:2004ne} and becomes
formally equal to the BFKL
equation~\cite{Kuraev:1977fs,Balitsky:1978ic}.

\subsection{Event Shape/Energy Flow Correlations}

An alternative to resumming the secondary logarithms
numerically or
analytically in the limit or large numbers of colors, is to
limit the radiation into the secondary part of phase space
by studying correlations between the non-global logarithms
and another global observable. Thus ``mini-jets'' of high
energy $\bar{Q}_{\bar{\O}} \rightarrow Q$ are suppressed,
because the radiation into $\bar{\O}$ is now restricted by
the value of the global observable. If the global observable
in addition has an adjustable parameter, such as the
family of angularities, Eq.~(\ref{eventdef}), the
importance
of the non-global components relative to the primary
energy flow can be controlled.

To be specific, we discuss the energy flow/event shape
correlations for dijet events in $e^+e^-$ annihilation
where color flow does not
add extra complications. We will discuss color flow in hadronic
collisions in the
next section.
We measure the energy flow into $\O$ by weighing the final state
$N$ with
\begin{equation}
\varepsilon(N) =  {1\over Q}\ \sum_{\hat n_i\in\O} Q_i\,
\label{eflowdef},
\end{equation}
where $Q_i$ denotes the energy of parton $i$ which is emitted into
the region of interest, $\O$. This is correlated with the event shape
(\ref{eventdef}), measured in $\bar{\O}$.
Thus the cross section is given by
\begin{equation}
{d \bar{\sigma}(\varepsilon,\tau_a,Q)\over d \varepsilon
\,d\tau_a}
=
{1\over 2Q^2}\ \sum_N\;
|M(N)|^2\,
\delta( \varepsilon-\varepsilon(N))\, \delta(\tau_a -\tau_a(N))
\label{crossdef}
\end{equation}
where we sum over all final states $N$ that contribute to the
weighted event, and where $M(N)$ denotes
the corresponding amplitude for ${\rm e^+e^-}\rightarrow N$.

\subsubsection{Resummation}

We are again interested in dijet events, which correspond to the
limit of small $\tau_a$ and $\varepsilon$, $\tau_a,\varepsilon \ll
1$. In this limit, as discussed above, the cross section receives
large logarithmic corrections in both variables. However, since
for the shape/flow correlation all of phase space is taken into
account, we can apply standard factorization and resummation
techniques~\cite{Collins:1989gx,Sterman:2004pd}.

We
arrive at the following factorized cross section in convolution form:
\begin{eqnarray}
 {1\over \sigma_{\rm tot}} \, \frac{d \bar{\sigma}
 (\varepsilon,\tau_a,Q)}{d \varepsilon \,d\tau_a} & = &
 H\left(\frac{Q}{\mu}\right)
 \int d\tau_{a\,s} \, \bar{S}
 \left(\frac{\varepsilon Q}{\mu},
 \frac{\varepsilon}{\tau_{a\,s}},\frac{\tau_{a\,s} Q}{\mu}\right) \nonumber \\
 & &  \times \, \prod_{i=1,2} \int d\tau_{a\,J_i} \, \bar{J}_i
 \left(\frac{\tau_{a\,J_i} Q}{\mu}\right)
\delta\left(\tau_a - \tau_{a\,s} -\tau_{a\,J_1} - \tau_{a\,J_2}\right)\, .\,\,
\label{conv}
\end{eqnarray}
Here, $\mu$ is the factorization scale which we take for simplicity
equal to the renomalization scale. $H$ is a short-distance
function, where all momenta are far off-shell, of order $Q$. This
 hard scattering function
is therefore independent of the shape/flow correlations, and only depends
on the energy (and implicitly on the scattering angles).
$\bar{S}$ is a subdiagram that describes soft radiation away from the two
jets $\bar{J}_i$, which contain all information about radiation
collinear to the primary partons, and are thus constrained only by the
event shape variable $\tau_a$. In contrast, $\bar{S}$ depends on
both, the event shape variable $\tau_a$, and
 the wide-angle radiation into $\O$ measured via $\varepsilon$.
In Laplace transform space the convolution (\ref{conv}) becomes
a simple product,
\begin{eqnarray}
\frac{d\sigma\left(\varepsilon,\nu_a,Q \right)}{d\varepsilon} & = &
 \int^\infty_0 d \tau_a\, {\rm e}^{\;
-\nu_a\,
\tau_a}\ {d
\bar{\sigma}(\varepsilon,\tau_a,Q) \over d\varepsilon\,d \tau_a }\, .
\label{trafo2} \\
{1\over \sigma_{\rm tot}}
\frac{d\sigma\left(\varepsilon,\nu_a,Q \right)}{d\varepsilon}
& = &
H\left(\frac{Q}{\mu}\right)
 S\left(\frac{\varepsilon Q}{\mu}, \varepsilon \nu_a,
 \frac{Q}{\nu_a \mu}\right)
 \prod_{i=1,2} J_i
 \left(\frac{Q}{\nu_a \mu}\right)
 \label{factor},
\end{eqnarray}
where now the unbarred quantities denote the Laplace transforms of
the corresponding functions in (\ref{conv}).

Resummation of the large logarithms in $\varepsilon$ and $\nu_a$
 is now straightforward.
We use the fact that the physical cross section is independent
of the factorization scale
\begin{equation}
\mu \frac{d}{d \mu} \left(\frac{d\sigma\left(\varepsilon,\nu_a,Q \right)}{d\varepsilon} \right)
= 0
\end{equation}
to obtain the following renormalization-group equations
for the soft function,
\begin{equation}
\mu \frac{d}{d\mu} \ln S \left(\frac{\varepsilon Q}{\mu}, \varepsilon \nu_a,
 \frac{Q}{\nu_a \mu}\right)
= - \gamma_s(\as(\mu)) \, ,
\label{RGE}
\end{equation}
where $\gamma_s$ is the soft anomalous
dimension. Analogous equations are found for the jet functions.
The anomalous dimensions can
depend only on variables held in common between  at least two
of the functions.  Because each function is infrared safe,
while ultraviolet divergences are present only in virtual
diagrams, the anomalous dimensions cannot depend on
the parameters $\nu_a$, $\varepsilon$ or $a$.  This leaves
as arguments of the $\gamma$s only
the  coupling $\as(\mu)$. The solution to Eq.~(\ref{RGE})
is found easily,
\begin{equation}
 S \left(\frac{\varepsilon Q}{\mu}, \varepsilon \nu_a,
 \frac{Q}{\nu_a \mu}\right) =
  S \left(\frac{\varepsilon Q}{\mu_0}, \varepsilon \nu_a,
 \frac{Q}{\nu_a \mu_0},a\right)
 e^{-\int\limits_{\mu_0}^\mu \frac{d \lambda}{\lambda}
 \gamma_s(\as(\lambda))} . \label{resum}
 \end{equation}
Choosing $\mu_0$ of order $\nu_a$ relegates all large logarithms
$\ln \nu_a$ into the exponent, as desired. By employing further
evolution equations and proper choices for the initial scales we
achieve exponentiation of all large logarithms $\ln \nu_a$ and
$\ln \varepsilon$, leaving only logarithms of order $\ln
(\varepsilon \nu_a)$ unexponentiated. For the full, a bit unwieldy
final resummed expression and all technical details, we refer to
Refs.~\cite{Berger:2003iw,Berger:2003zh}.

To summarize, large logarithms stemming from \emph{either} the
restriction of radiation by the global event shape \emph{or} by
the non-global observable can be resummed by standard techniques
into a Sudakov form analogous to Eq.~(\ref{thrustcomp}), leaving
only an unexponentiated piece from soft wide-angle radiation that
contributes logarithms of order $\ln (\varepsilon \nu_a)$, or
equivalently, upon transformation back from moment space, of order
$\ln (\varepsilon/\tau_a)$. Moreover, we observe that the not yet
fully exponentiated logarithms $\ln (\varepsilon/\tau_a)$ stem
from wide-angle soft radiation that completely decouples from the
jets. This soft radiation can be described by functions
constructed entirely out of non-abelian phases, or Wilson
(eikonal) lines~\cite{Collins:1989gx}. It is well-known that such
eikonal quantities exponentiate directly by a reordering of color
factors and with the help of an identity for eikonalized
propagators~\cite{Sterman:1981jc,Gatheral:1983cz}. Therefore all
large logarithms, including logarithms of $(\varepsilon/\tau_a)$
exponentiate, as was also found in Ref.~\cite{Dokshitzer:2003uw}
by an alternative analysis via the coherent branching formalism.
Furthermore, by choosing a parameter-dependent global event shape
we are able to dial the importance of these correlated logarithms
$\sim \ln (\varepsilon/\tau_a)$, and thus study the behavior of
the non-global logarithms.

\section{Color Flow}\label{color}

Above, we have discussed energy flow observables, which provide
information that is in some sense complementary to what we learn
from the study of event shapes. Event shapes shed light on the
global distribution of radiation, with the main features
determined by the underlying hard scattering. Long-distance
effects soften these main features, but not drastically so.
Interjet energy flow, on the other hand, reflects the interference
between radiation from different jets~\cite{Dokshitzer:1987nm},
and encodes the mechanisms that neutralize color in the
hadronization process. The study of the interplay between energy
and color flow in hadronic collisions~\cite{Kidonakis:1998nf} may
therefore help identify the underlying
event~\cite{Skands:2003yn,Acosta:2004wq}, to distinguish QCD
bremsstrahlung from signals of new physics.

\subsection{Hadronic Angularities}

In order to study color flow we need to
define event shape observables suitable for hadronic collisions.
Because parton-parton scattering is highly singular
in the forward direction, with a $1/\sin^4(\theta/2)$
behavior due to gluon exchange, $e^+e^-$ event shape definitions
 have to be adapted to accommodate this
 behavior  in the hadronic case. For further discussion of hadronic event shapes we
 refer to Ref.~\cite{Banfi:2004nk}.
 Again, the introduction of a parameter-dependent
 observable allows us to control the aforementioned difficulties.

We propose the following extension of the set of angularities to
hadronic collisions~\cite{Sterman:2005bf},
\begin{equation}
\tau_a(N) = \frac{1}{\sqrt{s}} \sum\limits_{i \in N}
\ p_{\perp,i }\;
e^{-|\eta_i|{ (1-a)}}\,   \label{hadeventdef}\, ,
\end{equation}
Here, in contrast to Eq.~(\ref{eventdef}), the transverse
momentum $p_{\perp,i}$ and rapidity $\eta_i$ of
parton $i$ is measured relative to the
collision axis in the c.m. frame. The parameter $a$
allows us to control the sensitivity to the forward
direction.

The hadronic cross section for the process
$A+B \rightarrow \mbox{ Jets } + X$ at c.m. energy $\sqrt{s}$
is now given by
a convolution of standard parton distribution functions (PDFs)
$\phi_{f/H}$ with a short-distance
partonic cross section $d \hat{\sigma}$:
\begin{eqnarray}
\frac{d\sigma_{AB}(\tau_a,p_\perp)}{d\eta dp_\perp  d \tau_a}
&= & \sum_{\rm f} \int dx_Adx_B \;
{\phi}_{f_A/A}\left(x_A, \mu_F\right) {\phi}_{f_B/B}
\left(x_B,\mu_F\right)\, \nonumber \\
& & \qquad \times \, \delta\left( p_\perp-{\sqrt{\hat s}\over
2\cosh\hat\eta}\right)\,
\frac{d \hat{\sigma}^{\rm (f)}(\tau_a,p_\perp,\mu_F)}
{d\hat\eta  d \tau_a}\, .
\label{hatsigfact}
\end{eqnarray}
$\mu_F$ denotes the factorization scale.
The PDFs $\phi_{f/H}$ describe the probability for finding
parton $f$ in hadron $H$ with momentum fraction $x$.
Hatted quantities are given in the partonic c.m. frame,
which can be found from the corresponding
hadronic quantities via $\hat\eta=\eta-(1/2)\ln(x_A/x_B)$, and
$\hat s=x_Ax_Bs$. The superscript $(\rm f)$
denotes the  Born-level $2 \rightarrow 2$ processes,
\begin{equation}
{\rm f}: \quad f_A + f_B \rightarrow f_1 + f_2, \label{pprocess}
\end{equation}
and the
sum is over all possible processes.
 The hard scale is set by
the transverse momentum of the observed jet, $p_\perp$.
Corrections to this leading-twist factorization
begin in general with powers of $\LQCD^2/p_\perp^2$ due to
multiple scatterings of partons.

As above, the partonic cross section will receive large
logarithmic corrections due to gluon radiation. However,
in the case of (\ref{hadeventdef}), $\tau_a \neq 0$ at
lowest order, due to the contributions of the outgoing
jets (recall that we measure with respect to the beam axis).
Thus, we encounter large logarithms of order
$\ln \tilde{\tau}_a$,
$\tilde{\tau}_a \equiv \tau_a -  \sum_{J=1,2} p_{\perp,J} e^{ - | \eta_J | (1-a)}$,
where the sum is over the outgoing jets.
In what follows, moments are taken with respect to $\tilde{\tau}_a$.

\subsection{Color Evolution}

The partonic cross section, $d \hat{\sigma}^{\rm (f)}/(d\hat\eta  d \tau_a)$
can be refactorized analogous to Eq.~(\ref{factor}), but now we have to take the
non-trivial color flow due to colored initial-state partons into account.
The regions that give leading contributions are as in the previous section a
hard scattering, soft, and jet functions, two for the outgoing jets,
and two jet functions for the beam jets.
The latter jets must be defined to avoid double counting due to the
parton distribution functions. Furthermore,
care must be taken in the definition of the outgoing jet functions due to
the logarithmic enhancement proportional to $\ln \tilde{\tau}_a$ instead of $\ln \tau_a$.
Such definitions are quite non-trivial,
and we do not attempt a full treatment here.
Instead, we sketch
the main features that emerge independently of the particular details
of the chosen observable.

\begin{figure}[htb]
\begin{center}
\epsfig{file=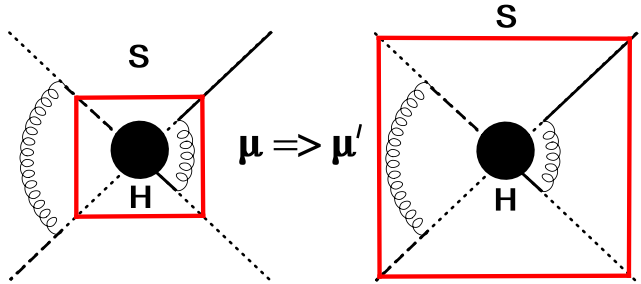,height=3.7cm,clip=0}
\vspace*{1mm}
\caption{Cartoon illustrating color evolution. $H$ and $S$ are the
hard scattering and the soft function, respectively, as described in the
text.} \label{figcolor}
\end{center}
\end{figure}

As was observed, for example, in
Refs.~\cite{Berger:2001ns,Kidonakis:1998nf,Sterman:2002qn}, there
is no unique way of defining color exchange in a finite amount of
time since gluons of any energy, including soft gluons, carry
octet color charge. The functions from which we construct the
refactorized partonic cross section are therefore described by
matrices in the space of possible color exchanges. This is because
as the (re)factorization scale changes, gluons that were included
in the hard function become soft and vice versa, as illustrated in
Fig.~\ref{figcolor}. Due to intrajet
coherence~\cite{Dokshitzer:1987nm}, however, the evolution of the
jets themselves is independent of the color exchanges. Once a jet
is formed, collinear radiation cannot change its color structure.
Therefore, the refactorized partonic shape/flow correlation can be
written in moment space as
\begin{equation}
    {d \hat{\sigma}^{\rm (f)}(\nu_a,p_\perp,\mu_F)\over d \hat{\eta}}
 =
H^{\rm (f)}_{LI}(p_\perp,\hat{\eta},{\mu},\mu_F)\;
S^{\rm (f)}_{IL}(\nu_a,\hat{\eta},{\mu}) \!\prod_{c=A,B,1,2}\!
J^{\rm (f)}_c(\nu_a,p_\perp,\hat{\eta},{\mu},\mu_F).
\label{trafosighad}
\end{equation}
analogous to the corresponding correlation in $e^+e^-$ events,
Eq.~(\ref{factor}). Now the hard and soft functions, $H$ and $S$,
respectively, are matrices in the space of color flow. Repeated
indices in color space, $L,I,$ are summed over. The superscripts
$(\rm f)$ label the underlying partonic process, as in
(\ref{pprocess}). For a list of convenient color bases for the
various $2 \rightarrow 2$ processes see for example
Refs.~\cite{Berger:2003zh,Kidonakis:1998nf}. $\mu$ is a
refactorization scale, not necessarily equal to the factorization
scale in Eq.~(\ref{hatsigfact}).

The color flow of the event shape is captured by the renormalization-group
equation (RGE) that results quite analogously to Eq.~(\ref{RGE}) from the
requirement that the physical cross section be independent of the
refactorization scale. Since $H$ and $S$ are matrices in color space,
the RGE is now a matrix equation, with anomalous dimension matrices $\Gamma$,
\begin{eqnarray}
\left(\mu\frac{\partial}{\partial\mu}+\beta(g_s)
\frac{\partial}{\partial g_s}\right) S_{IL}(\nu_a,\hat{\eta},{\mu})
& = &  -\left({\Gamma}^{\rm (f)}(\hat\eta,\as(\mu))\right)^{\dagger}_{IJ}
S_{JL}(\nu_a,\hat{\eta},{\mu}) \nonumber \\
& & -
S_{IJ}(\nu_a,\hat{\eta},{\mu}) \left({\Gamma}^{\rm (f)}(\hat\eta,\as(\mu))\right)_{JL}
\, .
\label{rgS0}
\end{eqnarray}
Since the anomalous dimension matrices for global event shapes are
found as usual from virtual graphs only, the matrices are
independent of the shape function and thus universal. They are
tabulated in various places, for example in
Refs.~\cite{Berger:2003zh,Kidonakis:1998nf}, and have been
implemented in the automated resummation program
CAESAR~\cite{Banfi:2004yd}.

For example, for the computation of the soft anomalous dimension matrix
for the partonic subprocess
$q_A \bar{q}_B \rightarrow q_1 \bar{q}_2$ the $t$-channel
singlet-octet basis is convenient,
\begin{eqnarray}
c_1 & = & \delta_{r_A,\,r_1} \delta_{r_B,\,r_2}, \nonumber \\
c_2 & = & - \frac{1}{2 N_c} \delta_{r_A,\,r_1} \delta_{r_B,\,r_2} + \frac{1}{2}
 \delta_{r_A,\,r_B} \delta_{r_1,\,r_2}, \label{qqbarbas} \, ,
\end{eqnarray}
where $r_i$ label the color indices of parton $i$, and $N_c$ is the number
of colors. The soft anomalous dimension matrix in this basis is given by,
\begin{equation}
{\Gamma}^{(q\bar{q}\rightarrow q \bar{q})} = \left( \begin{array}{cc}
2 C_F T &
- \frac{C_F}{N_c} U \\
-2 U \quad & - \frac{1}{N_c} (T- 2 U)
 \end{array}   \right),
\end{equation}
where $T$ and $U$ are functions of the Mandelstam
variables $\hat{s},\hat{t},\hat{u}$ in the partonic c.m. frame:
$T = \ln \left( \frac{-\hat{t}}{\hat{s}} \right) + i \pi,\,
U = \ln \left( \frac{-\hat{u}}{\hat{s}} \right) + i \pi$.

The partonic cross section resumming single logarithms in the color flow
is then found by solving
Eq.~(\ref{rgS0}),
\begin{eqnarray}
\frac{{d\hat{\sigma}}^{\rm (f)}
\left(\nu_a, p_\perp,\mu_F)\right) }
{d\hat\eta }
&=& \prod_{c=A,B,1,2}\!\!\!
J^{\rm (f)}_c(\nu_a,p_\perp,\hat{\eta},{\mu},\mu_F) \label{hadronLL}  \\
\times \sum_{\beta,\,\gamma}\; \!\!\!\!\!& & \!\!\!
          H^{\rm (f)}_{\beta\gamma}\left(p_\perp,\hat\eta,
\mu_F)\right) \,S^{\rm (f,0)}_{\gamma\beta}
\, e^{- \int\limits_{p_\perp/\nu_a}^{p_\perp}
\frac{d\lambda}{\lambda} \left[{\lambda}^{\rm (f)\star}_{\gamma}\left(\hat\eta,\as(\lambda)\right) +
{\lambda}^{\rm (f)}_{\beta}\left(\hat\eta,\as(\lambda)\right)\right]} \, , \nonumber
\end{eqnarray}
where we have transformed the matrices $H$ and $S$ to a basis, where the
anomalous dimension matrices $\Gamma$ are diagonal.
\begin{equation}
\left(\Gamma_{\mbox{\tiny eik}}^{\rm (f)}(\hat\eta)
\right)_{\gamma\beta} \equiv
\lambda^{\rm (f)}_\beta(\hat\eta)\delta_{\gamma\beta}=
R^{\rm (f)}_{\gamma I\, }\left(\Gamma_{\mbox{\tiny eik}}^{\rm
(f)}(\hat\eta)\right)_{IJ}\,
R^{\rm (f)}{}^{-1}_{J\beta},
\label{Gamdia}
\end{equation}
with $\lambda^{\rm (f)}_\beta$
the eigenvalues of $\Gamma$, and $R$ is the transformation matrix.
Greek indices $\beta, \; \gamma$ indicate that a matrix
is evaluated in the basis where
the eikonal anomalous dimension has been diagonalized.

From Eq.~(\ref{hadronLL}) we see that the color
flow at NLL is encoded in event shape independent
 anomalous dimension matrices. It may therefore be
 possible to find a suitable range
of the parameter $a$ in Eq. (\ref{hadeventdef})
or perhaps another, different, event shape observable,
where small-$x$ effects and problems due to incomplete
detector coverage in the forward direction
become unimportant, and the study of color flow
is unobstructed. Moreover, by combining the
analysis of color flow presented above with the
 study of energy flow discussed in the
previous section, as outlined in Ref.~\cite{Berger:2003zh}, we may
extract further important information about the dynamics of QCD.

\section{Conclusions and Outlook}

Above, we have illustrated how the introduction of a
parameter-dependence can further increase
the amount of information that can be extracted from the
study of event shape observables.
Aside from information about particle spin and various
parameters of QCD, such as color factors and the running coupling,
event shapes, especially their distributions,
 can be used to study the interplay of short-
and long-distance dynamics, and obtain insight into the
mechanisms of confinement, energy and color flow.

The study of the family of angularities is but a starting point.
Given the amount and versatility of event shape
observables for processes ranging from electron-positron annihilation
over deep inelastic scattering to hadronic collisions,
there are many more
possibilities and applications. It is certain that continued
theoretical and experimental efforts in this direction,
at present and future
accelerators, will reveal a wealth of information, and probably
 a few surprises.

\section*{Acknowledgments}

I thank Tibor K\'ucs, Lorenzo Magnea, and George Sterman for
fruitful collaborations whose results are described above. I also
thank George Sterman for comments on the manuscript.
Some of the illustrations above have been drawn with the help of
Jaxodraw~\cite{Binosi:2003yf}, using
Axodraw~\cite{Vermaseren:1994je}.

\end{document}